%% file: main.tex
\documentstyle[psfig]{mn}

\newcommand{\dd}{\mbox{d}}
\newcommand{\etal}{{\rm et~al.}}

\newcommand{\keV}{\hbox{keV}}
\newcommand{\Msun}{\hbox{$\rm\, M_{\odot}$}}
\newcommand{\TCDM}{\mbox{$\tau$CDM}}
\newcommand{\spose}[1]{\hbox to 0pt{#1\hss}}
\newcommand{\approxlt}{\mathrel{\spose{\lower 3pt\hbox{$\sim$}}
        \raise 2.0pt\hbox{$<$}}}
\newcommand{\approxgt}{\mathrel{\spose{\lower 3pt\hbox{$\sim$}}
        \raise 2.0pt\hbox{$>$}}}
\newcommand{\approxpropto}{\mathrel{\spose{\lower 3pt\hbox{$\sim$}}
        \raise 2.0pt\hbox{$\propto$}}}
\renewcommand{\vec}[1]{\mbox{\boldmath$#1$}}

\title[A simulated $\tau$CDM cosmology cluster catalogue]
{A simulated $\tau$CDM cosmology cluster catalogue: the NFW profile
and the temperature-mass scaling relations}
\author[P.~A.~Thomas \etal]{
  P.~A.~Thomas$^1$\thanks{Email: p.a.thomas@sussex.ac.uk},
  Orrarujee~Muanwong$^1$,
  F.~R.~Pearce$^3$,
  H.~M.~P.~Couchman$^2$,
\newauthor{A.~C.~Edge$^3$,
  A.~Jenkins$^3$ and
  L.~Onuora$^1$
  }\\
{}$^1$ Astronomy Centre, CPES, University of Sussex, Falmer, 
  Brighton, BN1\,9QJ\\
{}$^2$ Department of Physics \& Astronomy, McMaster University,
Hamilton, Ontario, L8S\,4M1, Canada\\
{}$^3$ Department of Physics, University of Durham, Science Laboratories,
  South Road, Durham DH1\,3LE
}

\pagerange{\pageref{firstpage}--\pageref{lastpage}}
\pubyear{2000}

\begin{document}

\maketitle

\label{firstpage}

\begin{abstract}

We have extracted over 400 clusters, covering more than 2 decades in
mass, from three simulations of the $\tau$CDM cosmology.  This
represents the largest, uniform catalogue of simulated clusters ever
produced.  The clusters exhibit a wide variety of density-profiles.
Only a minority are well-fit in their outer regions by the widely used
density profile of Navarro, Frenk \& White (1977, hereafter NFW).
Others have steeper outer density profiles, show sharp breaks in their
density profiles, or have significant substructure.  If we force a fit
to the NFW profile, then the best-fit concentrations decline with
increasing mass, but this is driven primarily by an increase in
substructure as one moves to higher masses.  The measured
temperature-mass relations for properties measured within a sphere
enclosing a fixed overdensity all follow the self-similar form,
$T\propto M^{2/3}$, however the normalisation is lower than in
observed clusters.  The temperature-mass relations for properties
measured within a fixed physical radius are significantly steeper then
this.  Both can be accurately predicted using the NFW model.

\end{abstract}


\input introduction

\input simulations

\input profiles

\input nfw

\input results

\input discussion

\input conclusions

\section*{Acknowledgments}

The simulations described in this paper were carried out on the
Cray-T3E at the Edinburgh Parallel Computing Centre as part of the
Virgo Consortium investigations of cosmological structure formation.
Interaction between authors was aided by a NATO Collaborative Research
Grant, CRG\,970081.  OM is supported by a DPST Scholarship from the
Thai government; PAT is a PPARC Lecturer Fellow; LO is a
Daphne-Jackson Fellow, funded by the Royal Society.

\input references

\label{lastpage}

\end{document}

%% file: introduction.tex
\section{Introduction}
\label{sec:intro}

Clusters of galaxies are used to constrain cosmological parameters as
they are the largest gravitationally bound systems and preserve
imprints of the evolution of the universe.  One of the most reliable
predictors for the mass-function of collapsed objects is that of Press
\& Schechter (1974).  Although based on na\"\i{}ve assumptions, it has
been widely used because of its simple form and, more importantly,
because the mass function is in excellent agreement with many N-body
simulations (e.g.~Lacey \& Cole 1993; but see Gross \etal~1998; Governato
\etal~1999; Jenkins \etal~2000).

In practice, cluster masses are not easily measured.  For this reason,
other properties such as X-ray luminosity or temperature are used to
estimate the mass.  The former of these is notoriously unreliable as
it depends strongly upon emission from the core of the cluster; hence
we will concentrate on the latter in this paper.  A simple application of the
virial theorem suggests that the mean temperature of an object that
has collapsed to the virial radius, $r_{\rm vir}$, follows the scaling
law
\begin{equation}
T_{\rm vir} \propto {M_{\rm vir}\over r_{\rm vir}}
\propto M_{\rm vir}^{2/3}(1+z_{\rm form}),
\label{eq:virial}
\end{equation}
where $M_{\rm vir}$ is the mass within $r_{\rm vir}$ and $z_{\rm
form}$ is the formation redshift.  In practice, clusters are assumed
to have formed at the redshift that they are observed, and the virial
radius is defined as the radius of a sphere enclosing an overdensity
of 180 (in critical-density cosmologies) relative to the mean density
at that redshift.  

Equation~\ref{eq:virial} is often assumed to be a rigorous theoretical
prediction, but this is not the case.  It assumes that the constant of
proportionality in Equation~\ref{eq:virial} is the same for low- and
high-mass clusters, but there is no reason why this need be true.  For
a power-law density fluctuation spectrum of dark-matter particles, the
theory of self-similarity tells us that the population of clusters at
one redshift is a scaled version of the population at another
redshift, but that is not the same as saying that coeval low- and
high-mass clusters are similar in form.  Even if the clusters do have
similar morphologies, we may still get different proportionality
constants if we measure their properties within the virial radius, as
this will correspond to different multiples of the half-mass radius in
each case.  

Another deficiency of Equation~\ref{eq:virial} is that it predicts
only the virial temperature---there is empirical evidence that the
observed (emission-weighted) X-ray temperature of the intracluster
medium, $T_X$, is greater than the virial temperature (e.g.~Edge \&
Stewart 1991a,b; Bahcall \& Lubin 1994).  Were these to be in constant
ratio, then the scaling relations would be preserved, however this is
unlikely to be the case.  Physical processes such as shock-heating and
radiative cooling act only on the gaseous component of clusters and
not on the dark matter.  Furthermore, these two processes have
different dependencies upon density and so will break the expected
self-similarity.  

Hjorth, Oukbir \& van Kampen (1998), in a sample of 8 clusters whose
masses are inferred from weak and strong gravitational lensing and
temperatures are measured from ASCA, confirm that (very roughly)
$T_X\propto M/r$.  Most observers, however, report the direct relation
between $T_X$ and $M$ without an explicit dependence upon $r$ and so
that is what we will do here.  The observations are discussed further
in later sections when we can compare them with our own results.

The scaling relation of Equation~\ref{eq:virial} has been tested using
hydrodynamical simulations with varying degrees of success.  We will
discuss these further in Section~\ref{sec:compsim}.  On the basis of
this, various authors have sought to constrain cosmological parameters
from the observed X-ray cluster temperature and luminosity functions,
for example, Lilje (1992), Viana \& Liddle (1996), Eke, Cole \& Frenk
(1996), Henry (1997), Markevitch (1998), Eke~\etal\ (1998), and Viana
\& Liddle (1999).  

Because the scaling laws of galaxy clusters are used so extensively as
the basis of cosmological models, we are undertaking a series of
N-body, hydrodynamical simulations to investigate the X-ray properties
of clusters of galaxies in detail.  It is intended eventually to
produce catalogues of clusters covering at least two decades in mass,
for several different variants of the CDM cosmology, and with a
variety of physical processes acting on the intracluster medium.  In
this current paper we report on the first of these catalogues for a
non-radiative simulation in the \TCDM\ cosmology.  The catalogue is
generated by combining three simulations (128$^3$ particles each of
gas and dark-matter) in which we resolve over 400 clusters at moderate
resolution.  Thus we have the correct boundary conditions for
large-scale structure and a reasonable dynamic range with which to
test the scaling relations.

We look at the circular velocity profiles of the clusters to see
whether they all have the same form, such as the commonly-used NFW
model.  We find both that individual clusters show a wide range of
profiles and that the parameters describing an average cluster are a
function of mass.  Thus, there is no a priori reason to suppose that
the cluster population should follow the scaling law described in
Equation~\ref{eq:virial}.  Nevertheless, it turns out, somewhat
fortuitously, that the deviations of the mean cluster population from
the scaling law are small, although the scatter from individual
clusters can be quite large.  When measured within a fixed physical
radius, however, the dependence of temperature upon mass is steeper
than that given by Equation~\ref{eq:virial}.

Because this current simulation does not include radiative cooling, it
cannot be an accurate reproduction of the real Universe.
Nevertheless, we have deliberately kept the model simple in order to
investigate how accurately, or otherwise, this simple model follows
the self-similar scaling laws between temperature and
mass\footnote{Because we are complete in mass, we quote $T$-$M$ rather
than $M$-$T$ relations}.  Future papers will then look at the
additional effects of cooling and non-gravitational heating.

We briefly describe the simulations and cluster identification in
Section~\ref{sec:numerics}, and also discuss our choice of softening
length.  In Section~\ref{sec:profiles}, we look at the density
profiles of the clusters and compare them with the NFW model.  We derive
predictions for the scaling relations in the NFW model in
Section~\ref{sec:nfw}, then compare these with those of our simulated
clusters in Section~\ref{sec:results}.  We compare our results with
previous work in Section~\ref{sec:discuss} and summarise our conclusions
in Section~\ref{sec:conclude}.

%% file: simulations.tex
\section{Numerical method}
\label{sec:numerics}

\subsection{The simulations}
\label{sec:sims}

We have carried out three simulations with 128$^3$ particles
each of gas and dark matter.  The cosmological parameters
were as follows: density parameter, $\Omega=1$; cosmological constant,
$\Lambda=0$; power spectrum shape parameter, $\Gamma=0.21$; and a
linearly-extrapolated root-mean-square dispersion of the density
fluctuations on a scale 8 ${h^{-1}}$ Mpc, ${\sigma_8}= 0.60$.  The
Hubble parameter is irrelevant as we did not allow the gas to cool;
the baryon fraction was set to a low value, $\Omega_b=0.06$ so that
the gas makes only a minor contribution to the gravitational
potential. The three simulations had different box-sizes,
corresponding to different mass-resolutions, as listed in
Table~\ref{tab:simpar}.  We also carried out two further simulations
of the middle-sized box to test the effect of changing the
gravitational softening: these are shown in italics in the Table.

\begin{table}
\caption{Run parameters for each of the simulations:
box size/$h^{-1}$Mpc; softening/$h^{-1}$kpc; dark-matter particle
mass/$h^{-1}M_\odot$; minimum resolved cluster mass/$h^{-1}M_\odot$;
minimum ratio of the 2-body relaxation time in the core of the
clusters to the age of the Universe.  The runs used in the
catalogues are shown in normal font; the comparison test runs are
shown in italics.}
\label{tab:simpar}
\begin{center}
\begin{tabular}{ccccc}
box& soft& $M_{\rm dm}$& $M_{\rm lim}$& $t_{r,{\rm min}}/t_0$ \\
\hline
\ 50.0&  20&  $1.6\times10^{10}$&  $8.22\times10^{12}$& 2.4\\
\it112.9& \it10& $\it1.8\times10^{11}$& $\it9.47\times10^{13}$& \it0.12\\
 112.9&  50&  $1.8\times10^{11}$&  $9.47\times10^{13}$& 3.0\\
\it112.9& \it100& $\it1.8\times10^{11}$& $\it9.47\times10^{13}$& \it11.9\\
 153.0& 68& $4.6\times10^{11}$& $2.37\times10^{14}$& 3.4
\end{tabular} 
\end{center}
\end{table}

We use a parallel version of the Hydra N-body/SPH code (Couchman,
Thomas \& Pearce, 1995; Pearce \& Couchman, 1997).  The simulations
were executed on the Cray T3E at the Edinburgh Parallel Computing
Centre as part of the Virgo Consortium's programme of investigations
into large-scale structure.

\subsection{Cluster identification}
\label{sec:clusid}

Initially we identify clusters in our simulation by searching for
groups of dark matter particles within an isodensity contour of 180,
as described in Thomas~\etal~(1998).  We work with a preliminary
catalogue of all objects with more than 30 particles, then retain only
those which have a total mass within the virial radius exceeding
$M_{\rm lim}$, corresponding to 500 particles of each species.  The
use of a small mass for the preliminary cluster selection ensures that
our catalogue is complete.  We have checked that using a different
isodensity threshold, a different selection algorithm, or using gas
particles instead of dark-matter particles to define the cluster,
leads to an almost identical cluster catalogue---the only difference
being the merger or otherwise of a small number of binary clusters.

We define the centre of the cluster to be the position of the densest
gas particle.  This will usually correspond to the peak of the X-ray
emission and has the advantage that it is independent of the cluster
selection method.

\subsection{Substructure statistic}
\label{sec:subst}

Much of the modelling that we will do on the clusters supposes that
they are smooth and spherically-symmetric.  In practice most clusters
show some degree of substructure.  Following Crone \etal~(1996) and
Thomas \etal~(1998), we measure substructure by comparing the
positions of the density maximum, $\vec{r}_{\rm d}$, and the centroid,
$\vec{r}_{\rm c}$ of the cluster, where the latter is averaged over
all particles within an isodensity contour of 180 times the
background density.  More specifically
\begin{equation}
\label{eq:subst}
S={|\vec{r}_{\rm d}-\vec{r}_{\rm c}|\over r_{180}},
\end{equation}
where $r_{180}$ is the radius of a sphere enclosing a mean density
equal to 180 times the background density.

In some of the figures that follow, we indicate the clusters with the
more prominent substructure, $S>0.2$, by plotting them with open
symbols: 14 per cent of the clusters fall into this category.

\subsection{Choice of softening}
\label{sec:soft}

In any N-body simulation, it is necessary to introduce a gravitational
softening in order that 2-body interactions do not become important.
Thomas \& Couchman (1992) estimated the minimum ratio of the 2-body
relaxation time to the age of the Universe (which occurs in the core
of the cluster, close to the softening radius) to be
\begin{equation}
  \label{eq:t2body}
  {t_{r,{\rm min}}\over t_0}
  \sim 1.3\,N_s^{1\over2}\left({{\rm soft}\over 10\,h^{-1}{\rm kpc}}
  {10\,h^{-1}{\rm Mpc}\over{\rm box}}\right)^{3\over2}
\end{equation}
where we have put in parameters appropriate to the current
simulations.  Here $N_s$ is the number of particles within $\sqrt{e}$
times the softening length. The values given in Table~\ref{tab:simpar}
are for the smallest clusters extracted from each run; the relaxation
time scales as $T^{1/2}$ for larger clusters because these have more
particles within the softening length.

In a similar calculation to that of Thomas \& Couchman, Steinmetz \&
White (1997) estimated the mean 2-body heating timescale for the cluster as
a whole.  They found that
\begin{equation}
\label{eq:sw}
\langle t_r\rangle \sim \sqrt{3\over\pi}\,{3N\over32\ln\Lambda}\,t_c,
\end{equation}
where $N/2$ is the number of particles within the half-mass radius,
$R_h$, of the halo, $\ln\Lambda\approx3$ is the Coulomb logarithm, and
$t_c=2\pi R_h/v_c$ is the orbital period at $R_h$, where $v_c$ is the
circular velocity.  The dependence upon the softening is much smaller
than in Equation~\ref{eq:t2body} which was measuring numerical
relaxation in the core of the cluster.  Putting in numbers appropriate
to our halos, we find the heating timescales for the lowest-mass
clusters in each run are approximately 10 times the age of the
Universe (the heating rate scales roughly in inverse proportion to
mass for higher-mass clusters).  Thus we might expect the gas to be
heated by up to 10 percent relative to the dark matter (the results of
the tests, reported below, suggest that the heating is slightly
smaller than this).

\begin{figure}
  \begin{center}
    \psfig{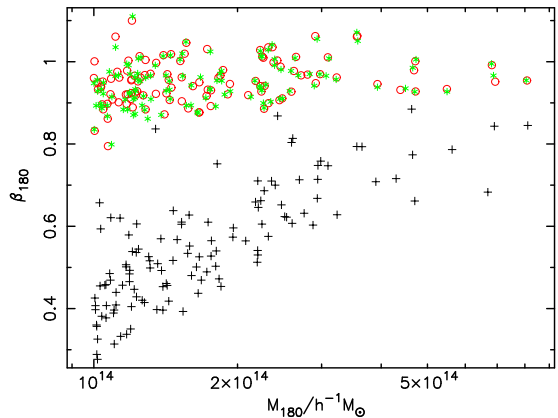}
    \caption{The ratio of the mean specific energy (thermal plus
      kinetic) of all particles to the gas specific energy within the virial
      radius for three different values of the softening: crosses,
      10\,$h^{-1}$kpc; circles, 50\,$h^{-1}$kpc; stars,
      100\,$h^{-1}$kpc.}
    \label{fig:betatest}
  \end{center}
\end{figure}

Because theoretical estimates of the numerical heating are so
uncertain, we tested the sensitivity of our results to changes in the
softening.  Three simulations of the 112.9\,$h^{-1}$Mpc box were
carried out which differed only in having softenings of 10, 50 and
100\,$h^{-1}$kpc, as shown in Table~\ref{tab:simpar}.  For each
cluster detected in the simulations, we measured the mean specific
energy of particles within the virial radius (the radius of a sphere
that encloses 180 times the mean density).  Figure~\ref{fig:betatest}
shows $\beta_{180}$, the ratio of the mean specific energy of all
particles to that of the gas particles.

As can be seen from the Figure, the effect of using too small a
softening can be severe.  Putting soft=$10\,h^{-1}$kpc into
Equation~\ref{eq:t2body} gives $t_{r,{\rm min}}/t_0\approx$0.12--0.24
for the lowest-to-highest mass clusters shown in the Figure.  The
Steinmetz \& White criterion makes only a very slight distinction
between the three runs with different softening (which enters
Equation~\ref{eq:sw} only via the Coulomb logarithm) and so should be
regarded as a necessary but not sufficient condition to prevent
artificial heating.

\begin{figure}
  \begin{center}
    \psfig{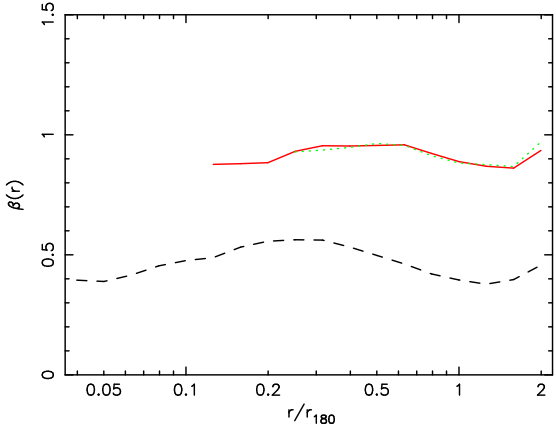}
    \caption{The ratio of the mean specific energy (thermal plus
      kinetic) to the gas specific energy as a function of radius
      for the 70-80 lowest-mass clusters that have minimal
      substructure, in the 112.9$\,h^{-1}$Mpc simulations.
      The dashed, solid and dotted lines are for softenings of 10, 50
      and 100$\,h^{-1}$kpc, respectively.}
    \label{fig:bprof}
  \end{center}
\end{figure}

Figure~\ref{fig:bprof} shows the profile of $\beta$ as a function
of radius (i.e.~$\beta(r)$ is the ratio of the mean specific energy of
all particles to the mean specific energy of gas particles within
spherical shells about the cluster centre).  Because individual
profiles show large variations in $\beta$, we have averaged the
profiles of the 70-80 smallest clusters,
$M<2\times10^{14}h^{-1}$\Msun, that show minimal substructure
($S<0.2$: see Section~\ref{sec:subst})---these are the clusters with
the lowest values of $t_{r,{\rm min}}/t_0$.  There is no evidence for
core heating in the runs with softenings of 50 and 100\,$h^{-1}$kpc,
whereas for the 10\,$h^{-1}$kpc softeing run $\beta$ is suppressed
 not only in the core but throughout the cluster.  This suggests that
much of the heating is going on in sub-halos before the cluster forms.
As the properties of these sub-halos are similar to those of the final
cluster core, this would explain why the Thomas \& Couchman formula
(Equation~\ref{eq:t2body}) is more appropriate than that of Steinmetz
\& White (Equation~\ref{eq:sw}).

It is clear from Figures~\ref{fig:betatest} and \ref{fig:bprof}, that
there is little difference between the results for softenings of 50
and 100\,$h^{-1}$kpc.  They have similar values of $\beta_{180}$ and
the radial variation in $\beta$ is minimal.  We conclude that a
softening of 50\,$h^{-1}$kpc is adequate for our purposes.

\begin{figure}
  \begin{center} \psfig{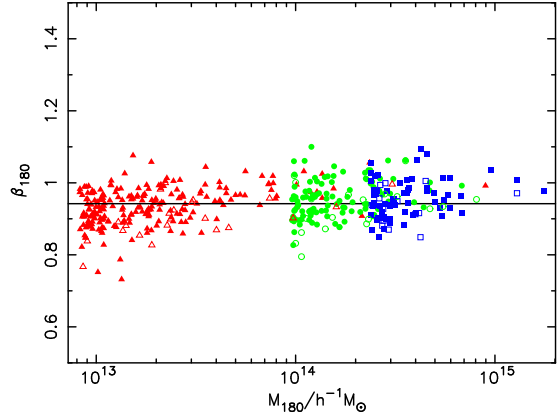}
    \caption{The ratio of the gas specific energy (thermal plus
    kinetic) to the dark matter specific energy within the virial
    radius for all clusters.  In this and subsequent figures, clusters
    extracted from the 50, 112.9 and 153\,$h^{-1}$Mpc boxes are plotted
    using triangular, circular and square symbols, respectively.  The 14
    percent of the clusters with $S>0.2$ are plotted with open
    symbols; the others with solid symbols.  The best fit constant is
    show by the solid line.}  \label{fig:beta180} \end{center}
\end{figure}

The softenings for each of the other two boxes were chosen to give a
similar central value of $t_{r,{\rm min}}/t_0$ for the lowest-mass
clusters as in the 50\,$h^{-1}$kpc softening run.  The values of
$\beta_{180}$ for each of the three production simulations are shown
in Figure~\ref{fig:beta180}.  Clusters with significant substructure,
$S>0.2$, are shown as open symbols; smooth clusters with solid
symbols.  In this figure and all those that follow, clusters from the
simulations of side 50.0, 112.9 and 153.0\,$h^{-1}$Mpc are shown as
triangular, circular and square symbols, respectively.

Navarro \& White (1993) and Pearce, Thomas \& Couchman (1994) have
shown that gas can gain energy from dark-matter during gravitational
collapse even in the absence of numerical heating.  The mechanism is
quite straight-forward: both the gas and dark matter are stirred
during the collapse, however the gas is able to thermalise its kinetic
energy, then pick up more at the expense of the dark matter.  There is
no reason to suppose that this process should have occurred equally in
low- and high-mass clusters.  There are hints of a slight decline in
$\beta_{180}$ at low masses in Figure~\ref{fig:beta180}, but 
to a good approximation, $\beta_{180}$ is independent of mass
and equal to 0.94$\pm0.03$.  Thus the fractional heating is comparable
to, but slightly smaller than, the estimate of 10 per cent that we
made from the Steinmetz \& White formula.

Because the gas has a greater specific energy than the dark matter, it
is more extended and hence the baryon fraction within the virial
radius is smaller than the global average.  We measure it to be
approximately $(0.85\pm0.02)\,\Omega_b$ in our simulated clusters,
independent of mass.

%% file: profiles.tex
\section{Cluster profiles}
\label{sec:profiles}

In clusters, the dark matter is dynamically dominant.  It is
therefore important to have a good model of the dark-matter mass
distribution.  Navarro, Frenk \& White (1995, 1996, 1997) showed that
the density profiles of simulated clusters in a wide variety of
cosmological models are well-described by the formula
\begin{equation}
\rho={\rho_0\over x\,(1+x)^2},
\label{eq:dens}
\end{equation}
where $r=ax$ is the radius and $\rho_0$ and $a$ are free parameters,
and went so far as to describe this as a `universal density profile'.
This formula must break down at large radii as it predicts infinite
mass, but it appears to hold out to the virial radius, defined (in
critical-density models) as the radius, $r_{180}$, enclosing a mean
overdensity of 180.

In a previous paper Thomas \etal\ (1998) found that the NFW formula
was indeed a good approximation to the mass distribution of an average
cluster, but that there was a wide dispersion in the rate at which the
density was declining at the virial radius.  Hence we introduce a more
general profile
\begin{equation}
\rho={\rho_0\over x\,(1+x)^s},
\label{eq:denss}
\end{equation}
where $s$ is a constant.  Because the name is now so well-established
in the literature, we call this the `generalised NFW' model, although
we note that the case $s=3$ was introduced first, by Hernquist (1990).

The resolution of the simulations presented in this paper is not
sufficient to determine the density profile in the cores of the
clusters.  This has been looked at in depth by Moore \etal~(1998) and
by Jing \& Suto (2000).  There is now evidence to suggest that the 
inner density cusp rises more steeply than $\rho\propto r^{-1}$, but
that is of little consequence for the overall dynamics of the cluster
and so we will stick with Equation~\ref{eq:denss} here.

\subsection{Best-fit circular speed profiles}
\label{sec:profvcir}

The density profiles of individual clusters are often very noisy which
makes them difficult to match to any given theoretical profile.  It is
better to use a cumulative profile such as the circular speed, $v_{\rm
c}$, which is anyway a dynamically more relevant quantity.  For the
above distribution,
\begin{equation}
v_{\rm c}^2\propto {(1+x)^{s-1}-(s-1)\,x-1\over x\,(1+x)^{s-1}}
\label{eq:vcir}
\end{equation}
when $s\neq2$, and
\begin{equation}
v_{\rm c}^2\propto {(1+x)\ln(1+x)-x\over x\,(1+x)}
\label{eq:vcir2}
\end{equation}
when $s=2$.

\begin{figure}
  \begin{center}
    \psfig{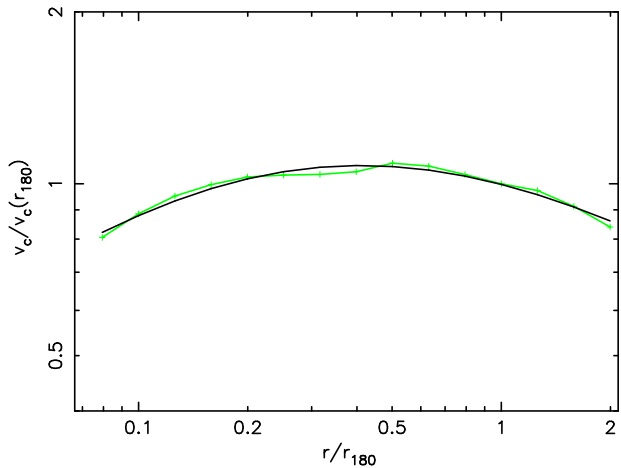}
    \caption{The circular-speed profile for one of the clusters.
      The best-fit theoretical model is shown by the smooth curve.}
    \label{fig:vcir}
  \end{center}
\end{figure}

An example of a circular-speed profile is given in
Figure~\ref{fig:vcir} together with the best-fit function of the form
of Equation~\ref{eq:vcir}.  We have fitted and plotted the curve
between twice the softening and twice the virial radius.  This
particular cluster was chosen because the goodness-of-fit as measured
by the mean-square deviation from the theoretical curve is the median
value for all the clusters.  As can be seen, there are kinks in the
profile, showing evidence of substructure: this is typical of most of
the clusters in our catalogue.

\begin{figure}
  \begin{center}
    \psfig{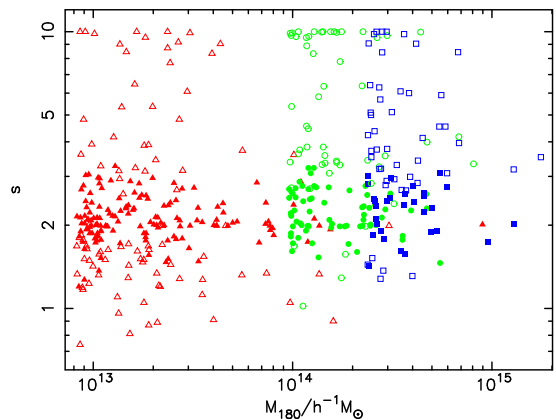}
    \caption{$s$ versus $M_{180}$ for all the clusters.  
    The solid symbols represent clusters for which the slope of the
    density profile at two virial radii is within 25 per cent of the
    asymptotic slope at large radii.}
    \label{fig:sm180}
  \end{center}
\end{figure}

The best-fit values of $s$ as a function of cluster mass are shown in
Figure~\ref{fig:sm180}.  Note that $s$ has been limited to be less
than or equal to 10.  
The asymptotic density profile at large radii has a slope of $-(s+1)$.
However, this is not always representative of the slope at two virial
radii, the outer radius to which we fit the rotation curve.  Hence we
plot with solid symbols in the Figure only those clusters for which
the two differ by less than $0.2s$---this corresponds to a
characteristic radius $a<0.5r_{180}$. 

A quick glance at Figure~\ref{fig:sm180} makes the suggestion that
$s=2$ is a universal density profile seem surprising.  However, it is
hard to measure the density profiles in the outer parts of clusters
with any degree of accuracy and the answer that one gets often depends
upon the radial extent of the fit.  At one virial radius, the slope of
the density profile is far from $-(s+1)$ and so the asymptotic slope
is poorly constained.  In addition, $s$ and $a$ are strongly
correlated and it is often possible to get a reasonable fit by forcing
$s=2$ and allowing $a$ to vary.  Hence the statement that the profile
within one virial radius is `consistent with and NFW profile' is
largely meaningless.  It is for this reason that we choose to fit the
profile within two virial radii instead.

We define clusters to be consistent with an NFW profile if the
asymptotic slope of their density profile lies between $-2.8$ and
$-3.2$ (i.e. $1.8<s<2.2$).  Just under a quarter of the clusters meet
this criterion.

\begin{figure}
  \begin{center}
    \psfig{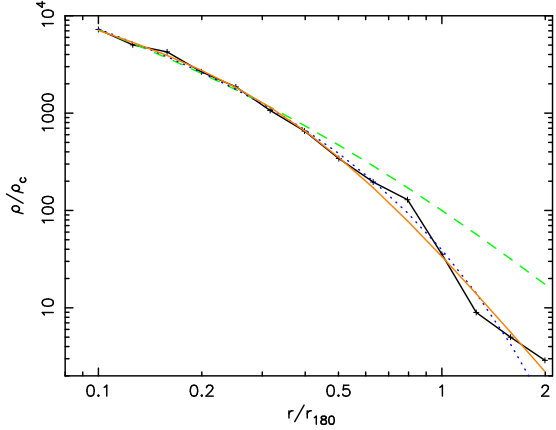}
    \caption{The density profile for one of the clusters.
      The dashed line shows the best-fit NFW model.  The dotted line
      shows the best-fit generalised NFW model with $s=10$.  The
      solid line shows a better, alternative model, as described in
      the text.}
    \label{fig:dprof}
  \end{center}
\end{figure}

The best-fit profiles of many clusters plotted with open symbols show
a high value of $s$.  However, this does not indicate steep density
profiles at large radii because the best-fit core radii rise to
compensate.  Rather, it indicates that the functional form of the
generalised NFW profile is a poor representation of the cluster.  As an
example consider the cluster shown in Figure~\ref{fig:dprof}.
This is a smooth cluster: visually it appears spherically-symmetric
and its substructure parameter is small, $S=0.038$.  In addition, the
best-fitting ellipsoid (see Thomas \etal~1998) is amongst the most
spherical of any cluster with axial ratios of 1.14:1.0:0.94.
However, the density profile, even out to 0.8 virial radii, is poorly
fit by an NFW profile (dashed line).  Also, it shows a sharp change in
slope at this radius that cannot be matched by any generalised NFW
profile.  The dotted line shows a generalised NFW model with $s=10.$ and
$a=3.1\,r_{180}$.  In order to reproduce the sharp decline in density
at the virial radius, $s$ has to be very large, but this then leads to
a density profile that is declining far too rapidly in the outer parts
of the cluster (and would steepen even more at radii larger than those
shown in the Figure).  A better representation of the density profile
in this case is given by the dotted line which corresponds to the
function
\begin{equation}
\rho={\rho_0\over x\,(1+x^2)^{s/2}},
\label{eq:densking}
\end{equation}
where $x=r/a$ as before, and $s=3.2$, $a=0.40\,r_{180}$.  This value of
$s$ is a much better estimate of the asymptotic slope of the density
profile at large radii.

It can be seen from Figure~\ref{fig:sm180} that the line $s=3$ roughly
separates the solid from the open symbols in the upper half of the
plot.  Thus, where the outer slope of the density profile is
well-defined, it generally lies between $-3$ (an NFW profile) and $-4$
(a Hernquist profile).  The open symbols represent clusters, like that
shown in Figure~\ref{fig:dprof}, that have a sharper break in their
density profile than can be fit by a generalised NFW profile: these comprise
about 30 per cent of the total cluster sample.

The open symbols that correspond to values of $s$ less than 2 are
mostly clusters that show some degree of substructure, for which the
density profile is not well-defined.  These comprise another 16
percent of the cluster population.

Although the spread in $s$ is large, there is a weak trend for $s$ to
increase with mass.  To make this more evident, we define an average
low-mass and an average high-mass cluster by selecting all relatively
smooth clusters, $S<0.2$, in the mass ranges
$M<1.2\times10^{13}h^{-1}$\Msun and $M>3.0\times10^{14}h^{-1}$\Msun.
The resulting profiles are extremely well-fit by our theoretical
model with slopes of $s=2.0$ and $s=2.3$, respectively.

It is clear from the above analysis that there is no `universal
profile' for dark matter halos.  A substantial proportion of clusters
show obvious substructure, and even those that don't exhibit a wide
variety of functional forms for the spherically-averaged density
profiles of halos.

\subsection{Best-fit NFW profiles and concentrations}
\label{sec:profnfw}

The concept of a universal density profile is an attractive one.  It
makes modelling of observed clusters much simpler and it has the
advantage that there is only one free parameter---the ratio of the
characteristic radius in the NFW formula to the virial radius,
$x_{180}=r_{180}/a(s=2)$.\footnote{NFW define $x_{200}$ to be the
`concentration parameter', presumably using 200 as an approximation
for the virial overdensity; in this paper we will use the term to
stand for $x_{180}$ instead---there is little difference between the
two.}  Therefore, we wish to see how well one can approximate cluster
properties by assuming that they all follow the NFW profile, in
defiance of the results of Section~\ref{sec:profvcir}.

\begin{figure}
  \begin{center} \psfig{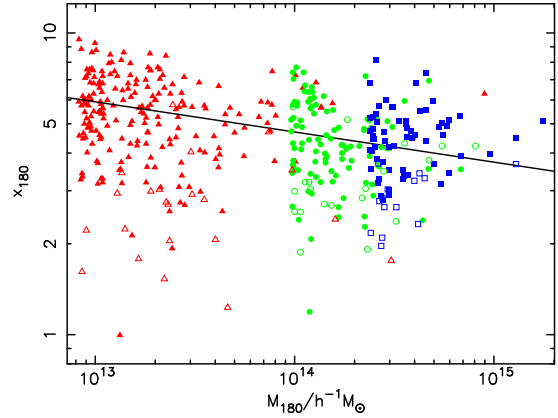}
    \caption{The measured concentration assuming an NFW profile,
    $x_{180}$, versus mass within the virial radius, $M_{180}$, for
    all the clusters.  The 14 percent of the clusters with $S>0.2$ are
    plotted with open symbols; the others with solid symbols.}
    \label{fig:xm180} \end{center}
\end{figure}

The best-fit values of $x_{180}$ as a function of mass are shown in
Figure~\ref{fig:xm180}.  There is a general trend of decreasing
concentration as one moves to higher masses, but once again the
scatter is large.  The solid line in Figure~\ref{fig:xm180} corresponds
to the function
\begin{equation}
x_{180}=3.7\,\left(M\over10^{15}h^{-1}\Msun\right)^{-0.1}.
\label{eq:xm180}
\end{equation}
The 86 per cent of the clusters with $S<0.2$ are spread equally above
and below the line.  We shall use this relation in the analysis that
follows to see how well the simple NFW model predicts the measured
scaling relations between temperature and mass.

We can compare our concentration parameters to those of Navarro, Frenk
\& White (1997), by measuring halo mass in terms of $M_*$, the mass of
a spherical region with root-mean-square linear density fluctuation of
1.69 today.  For our simulations, $M_*\approx
3.3\times10^{12}h^{-1}\Msun$, and $M_{180}/M_*$ ranges from 3 to 300.
This gives concentrations in good agreement with their CDM models.

\begin{figure}
  \begin{center} 
    \psfig{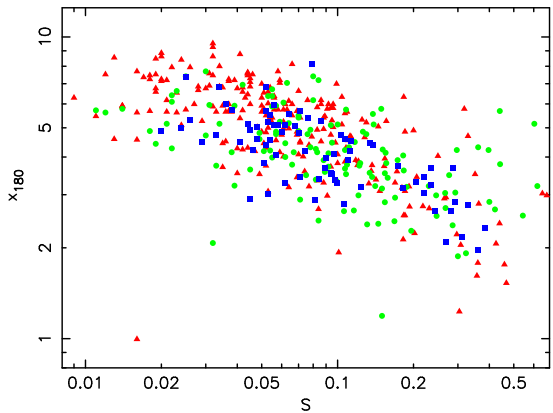}
    \caption{$x_{180}$ versus $S$ for all the clusters.  Note that six
    points to the left and the bottom of this plot have been omitted
    for clarity.}
    \label{fig:x180S} 
  \end{center}
\end{figure}

\begin{figure}
  \begin{center}
    \psfig{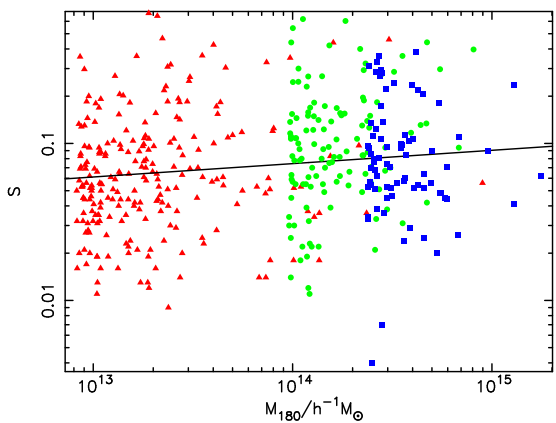}
    \caption{$S$ versus $M_{180}$ for all the clusters.  A couple of
    points have been omitted from the bottom of the plot, for clarity.
    The solid line shows a median line that divides both the triangles
    and the circles plus squares into two equal groups.}
    \label{fig:Sm180}
  \end{center}
\end{figure}

Interestingly, Figure~\ref{fig:x180S} shows that the measured
concentrations are highly-correlated with the substructure statistic.
Thus low concentrations do not normally occur in
spherically-symmetric clusters, but come from spherical averaging of
clusters that have substructure.  This suggests that the weak trend of
decreasing concentration with increasing mass is driven primarily by
the fact that substructure increases with mass, as shown in
Figure~\ref{fig:Sm180}.

\begin{figure}
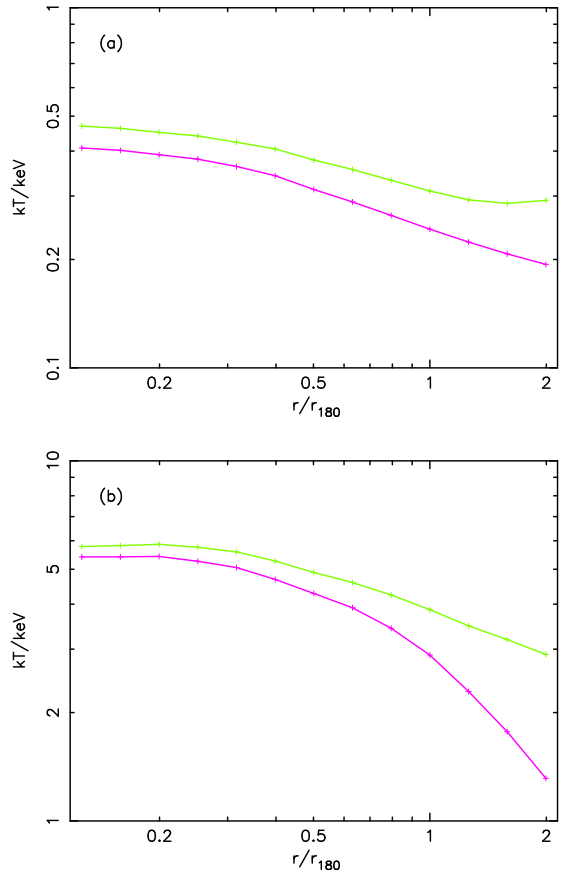

  \begin{center}
    \psfig{file=tprofa.cps,angle=270,width=8.1cm}
    \psfig{file=tprofb.cps,angle=270,width=8.1cm}
    \caption{Mean temperature profiles for (a) 79 low-mass, and (b) 34
    high-mass clusters, as described in the text.  The upper curve in
    each case shows the total specific energy of the gas and the lower
    curve the temperature, both in units of keV/k.}
    \label{fig:tmprof}
  \end{center}
\end{figure}

We conclude this section by plotting in Figure~\ref{fig:tmprof} the
temperature profiles for the mean clusters described above.  The lower
curve in each panel shows the gas temperature in units of keV$/k$,
while the upper curve shows the total specific energy of the gas
(thermal plus kinetic) in the same units.  It should be noted that the
gas is approximately isothermal within about one-third of the virial
radius, but that its temperature drops rapidly in the outer parts of
the cluster.  This is a reflection of the fact that the specific
energy of the dark matter declines in the outer regions of the
cluster, because the ratio between the two is approximately constant
out to two virial radii.

It is a matter of some debate in the literature as to whether the
observed temperature profiles of clusters are isothermal or decline to
large radii.  Using ASCA data, Irwin \& Bregman (2000) and White
(2000) infer flat or slowly increasing temperature profiles within
0.3\,$r_{\rm 180}$, whereas Markevitch \etal~(1998) find that the
temperature drops by a factor of two within 0.5\,$r_{180}$.
It is clear from our simulations that isothermal profiles beyond about 
0.3\,$r_{\rm 180}$ are inconsistent with our simple model and would
probably require some form of non-gravitational heating.

%% file: nfw.tex
\section{Predicted scaling relations}
\label{sec:nfw}

In this section, we predict scaling relations for model clusters under
the assumptions that they are spherically-symmetric, isolated and in
hydrodynamical equilibrium.  It is not that we are asserting that
these assumptions are true, but rather we want to see how well such a
simple model will perform and to test the sensitivity of the scaling
relations to changes in the model parameters.

By scaling everything in terms of a characteristic radius, $a$, and
density $\rho_0$, for each cluster, we can make all the variables
dimensionless.  We write $r=ax$ and $\rho=\rho_0\tilde\rho$, where
\begin{equation}
\tilde\rho={1\over x\,(1+x)^s}.
\label{eq:nfwdens}
\end{equation}
and the tilde indicates a dimensionless quantity.
Similarly, for the mass within radius $r$,
\begin{equation}
m=4\pi\rho_0\,a^3\tilde{m}
\label{eq:nfwmzero}
\end{equation}
where
\begin{equation}
\tilde{m}=\int_0^{r/a}\tilde\rho\,x^2\dd x.
\label{eq:nfwmass}
\end{equation}

When we observe clusters, we usually measure their properties at or
within the virial radius, $r_{\rm vir}=a\,x_{\Delta_c}$ at which their
enclosed density equals some multiple, $\Delta_c$, of the critical
density of the universe.  For the \TCDM\ model that we consider in this
paper, $\Delta_c=180$, but we will set $r_\delta=a\,x_{\delta}$ and
leave $\delta$ as a free parameter, so as to be completely general
in our argument.

From the above definitions, we can write $\rho_0$ in terms
of $x_\delta$,
\begin{equation}
\rho_0=60\,\delta_{180}\,\rho_c\,{x_\delta^3\over\tilde{m}_\delta},
\label{eq:nfwrhozero}
\end{equation}
where $\rho_c$ is the critical density, $\delta_{180}=\delta/180$ and
$\tilde{m}_\delta=\tilde{m}(x_\delta)$.  

\subsection{The virial temperature-mass relation}

We next calculate the velocity dispersion, $\sigma$ as a function of
radius for the NFW profile.  To do this, we assume that the velocity
ellipsoid is isotropic.  It then becomes convenient to write
$\sigma^2=kT/\mu m_H$, where $k$ is the Boltzmann constant, $\mu m_H$
is the mean particle mass for the gas, and $T$ is the `dynamical
temperature' of the halo.  If the specific energy of the gas and
dark-matter were the same, then $T$ would be equal to the gas
temperature.  

The spherically-symmetric Jeans' Equation for a fluid
with an isotropic velocity dispersion is
\begin{equation}
{\dd(\tilde{\rho}\tilde{T})\over\dd x}=
   -{\tilde{m}\tilde{\rho}\over x^2},
\label{eq:nfwtemp}
\end{equation}
where
\begin{equation}
T={G\mu m_H\over k}\,4\pi\rho_0a^2\,\tilde{T}.
\label{eq:nfwtzero}
\end{equation}

We are now in a position to calculate the mean dynamical temperature,
$T_\delta$, within radius $r_\delta$.  By definition,
\begin{equation}
\tilde{m}_\delta\,\tilde{T}_\delta=
\int_0^{\tilde{m}_\delta}\tilde{T}\,\dd\tilde{m}=
\int_0^{x_\delta}\tilde\rho\,\tilde{T}\,x^2\dd{x}.
\label{eq:nfwtempdelta}
\end{equation}

Combining Equations~\ref{eq:nfwmzero} and \ref{eq:nfwtzero}
to eliminate $a$, and then substituting for
$\rho_0$ from Equation~\ref{eq:nfwrhozero}, gives
\begin{eqnarray}
k\,T_\delta &=& G\mu m_H\,(240\pi\rho_c\,\delta_{180})^{1\over3}\,
              {x_\delta\,\tilde{T}_\delta\over\tilde{m}_\delta}\,
              m_\delta^{2\over3} \nonumber\\
            &\approx& 16.0\,\delta_{180}^{1\over3}\,
              {x_\delta\,\tilde{T}_\delta\over\tilde{m}_\delta}\,
              \left(m_\delta\over10^{15}h^{-1}\Msun\right)^{2\over3}\keV,
\label{eq:nfwtmvir}
\end{eqnarray}
where we have taken $\mu m_H\approx 10^{-27}$kg, corresponding to a
fully-ionized cosmic mix of elements.  

\begin{figure}
\begin{center}
 \psfig{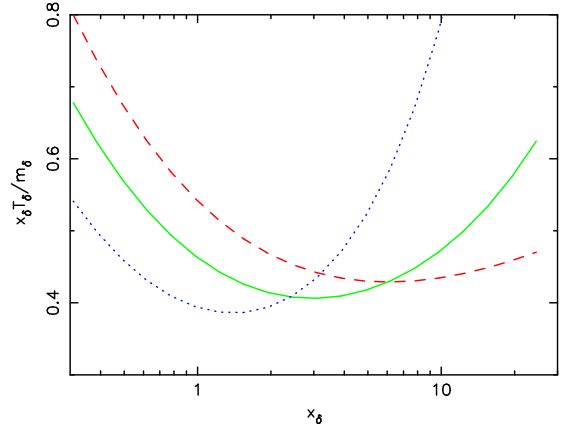}
 \caption{The shape factor
 $x_\delta\,\tilde{T}_\delta/\tilde{m}_\delta$ in the 
 dynamical temperature-mass relation of Equation~\ref{eq:nfwtmvir}.
 The dashed, solid and dotted lines correspond to $s=1.5$, 2 and 3,
 respectively.}
\label{fig:nfwxtom}
\end{center}
\end{figure}

When $\delta=\Delta_c$, then $T_\delta$ is the `virial temperature'.
We will use the notation $T_{180}$ rather than $T_{\rm vir}$, to make
explicit that we are referring to an overdensity of 180 (as in other
cosmologies, the virial radius would correspond to a different overdensity).

If all clusters are self-similar (e.g.~if they all follow the NFW
model), and if we measure the total mass, $m_\delta$, and mean
dynamical temperature, $T_\delta$, within $r_\delta$, then it is often
assumed that they must obey the relation $T_\delta\propto
m_\delta^{2/3}$.  This is not strictly true, however, because of the
presence of the $x_\delta \tilde{T}_\delta/\tilde{m}_\delta$ term
which varies with concentration, $x_\delta$.

We show the variation of $x_\delta\,\tilde{T}_\delta/\tilde{m}_\delta$
with $x_\delta$ in Figure~\ref{fig:nfwxtom} for values of $s=1.5$, 2
and 3.  We see that for values of $x_{180}$ in the range 1.5 to 10
that are typical when fitting the NFW profile to clusters,
$x_{180}\,\tilde{T}_{180}/\tilde{m}_{180}$ lies in the narrow range
0.4--0.47.  Fitting $s=1.5$ profiles instead would give smaller values
of $x_{180}$, and fitting $s=3$ profiles larger values, so that these
too give approximately the same answer.  Thus the virial
temperature-mass relation is robust to variations in the density profile.

\subsection{The gas density profile and the X-ray temperature-mass relation}

The virial temperature-mass relation predicted above is of little
use observationally.  Instead we need to use the emission-weighted,
X-ray temperature.
Previous studies such as Eke, Navarro \& Frenk (1998) and Makino,
Susaki \& Suto (1998) have calculated the mean X-ray temperature and the
luminosity of the intracluster medium assuming that it is isothermal
and sitting in an NFW potential.  Figure~\ref{fig:tmprof} shows that
the gas is far from isothermal and so we instead make the assumption,
as indicated by Figure~\ref{fig:bprof}, that the ratio of the
specific energies of the gas and dark matter is everywhere constant
and equal to $\beta=0.94$, i.e.\ $T_{\rm gas}=T/\beta$.

We will further assume that the dark matter is
dynamically dominant.  Then
\begin{equation}
{1\over\tilde{\rho}_{\rm gas}}\,
{\dd(\tilde{\rho}_{\rm gas}\tilde{T}_{\rm gas})\over \dd x}
={1\over\tilde{\rho}}\,{\dd(\tilde{\rho}\,\tilde{T})\over \dd x},
\label{eq:nfwtgastdyn}
\end{equation}
which can be expanded to give
\begin{equation}
{\dd\ln\tilde{\rho}_{\rm gas}\over\dd\ln x}
={\dd\ln\tilde{\rho}\over\dd\ln x}-(1-\beta)\,{\dd\ln\tilde{P}\over\dd\ln x},
\end{equation}
where $\tilde{P}=\tilde{\rho}\tilde{T}$.  This shows that the gas is
able to support itself more effectively and hence has a shallower
density gradient than the dark matter.  For $\beta=0.94$, the
asymptotic density gradient at large radii is $-2.7$, compared to $-3$
for the dark matter.

The mean X-ray temperature within radius $r_\delta$ is defined by
\begin{equation}
\tilde{L}_\delta\,\tilde{T}_{X,\delta}=
\int_0^{\tilde{L}_\delta}\tilde{T}_{\rm gas}\,\dd\tilde{L},
\label{eq:nfwtempxdelta}
\end{equation}
where
\begin{equation}
\tilde{L}(x)=\int_0^x\tilde{\rho}_{\rm gas}^2\,\tilde{T}_{\rm
gas}^{1\over2}\,\dd x
\label{eq:xlum}
\end{equation}
and $\tilde{L}_\delta=\tilde{L}(x_\delta)$.  Note that we have assumed
that the emissivity of the gas scales as $\tilde{T}_{\rm gas}^{1/2}$,
although this is only really true at high temperatures where
bremsstrahlung dominates.  However, the dominant contribution to the
variation in the integrand in Equation~\ref{eq:xlum} comes from
$\tilde{\rho}_{\rm gas}$ and so this is a reasonable approximation.

\begin{figure}
\begin{center}
 \psfig{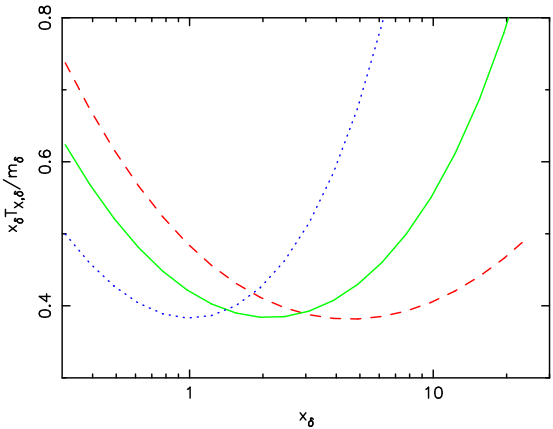} 
 \caption{The shape factor
 $x_\delta\,\tilde{T}_{X,\delta}/\tilde{m}_\delta$ in the X-ray
 temperature-mass relation.  The dashed, solid and dotted lines
 correspond to $s=1.5$, 2 and 3, respectively.}
\label{fig:nfwxtxom}
\end{center}
\end{figure}

The values of the normalisation factor in the X-ray temperature-mass
relation, $x_\delta\,\tilde{T}_{X,\delta}/\tilde{m}_\delta$, are shown
in Figure~\ref{fig:nfwxtxom}.  They can be seen to rise more steeply
at large concentrations than in Figure~\ref{fig:nfwxtom}.  This is
because the X-ray temperature declines far less steeply with radius
than the dynamical one used previously.  Figure~\ref{fig:nfwxtxom}
suggests that the range of concentrations seen in our simulated
clusters will add quite a lot of scatter to the mean X-ray
temperatures within the virial radius, with the more concentrated
clusters being hotter than the less-concentrated ones for a given
mass by up to 40 per cent.

%% file: results.tex
\section{Measured scaling relations}
\label{sec:results}

\subsection{The virial temperature-mass relation}
\label{sec:tmvir}

\begin{figure}
  \begin{center} \psfig{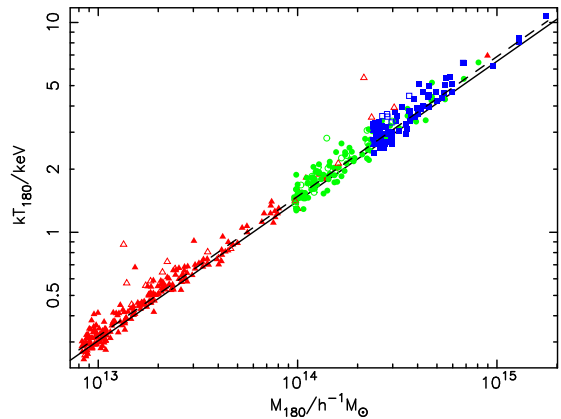}
    \caption{$T_{180}$ versus $M_{180}$ for all the clusters.  In this
    and subsequent figures in this section, the 14 per cent of
    clusters that have a substructure statistic $S>0.2$ are plotted
    with open symbols; others as filled symbols.  The dashed
    line shows the best-fit power-law.  The solid line shows the
    prediction from the isotropic NFW model.}  
    \label{fig:tm180}
    \end{center}
\end{figure}

The location of each of our clusters on the $T_{180}$-$m_{180}$ plane
is shown in Figure~\ref{fig:tm180}.  Also shown by the dashed line is
the best-fit power-law
\begin{equation}
k\,T_{180} \approx 6.9\,
              \left(m_{180}\over10^{15}h^{-1}\Msun\right)^{0.67\pm0.02}\keV,
\label{eq:tm180}
\end{equation}
which corresponds to a value of
$x_\delta\tilde{T}_\delta/\tilde{m}_\delta=0.43$.

The predicted relation from Equation~\ref{eq:nfwtmvir}, using the median
concentrations, $x_{180}$, taken from Equation~\ref{eq:xm180}, is shown
by the solid line in the Figure.  It can be seen that this predicts
temperatures that are slightly too low, although the spread in
temperatures is consistent with the spread in $x_{180}$ seen in
Figure~\ref{fig:xm180}.  

It would be possible to reconcile the predicted and measured relations
by using slightly larger values of $x_{180}$: Figure~\ref{fig:x180S}
shows that the most regular clusters are biased towards higher
concentrations.  An alternative explanation for the low virial
temperatures is that the dark matter velocity ellipsoid is not
isotropic, as radial orbits are less efficient at supporting the
particles within a given potential than transverse ones.  This is
discussed further in Section~\ref{sec:nfwani}.

\subsection{X-ray versus virial temperatures}

\begin{figure}
  \begin{center} \psfig{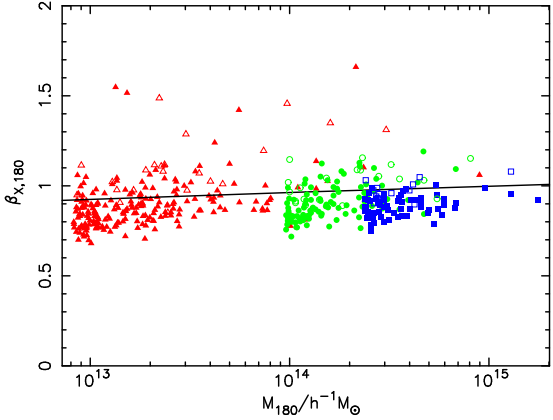}
    \caption{The ratio of the dynamical temperature to the X-ray
    temperature within the virial radius for all clusters.  The
    prediction from the isotropic NFW model is show by the solid
    line.}  \label{fig:betax180} \end{center}
\end{figure}

The ratios of the dynamical temperatures to the X-ray temperatures,
averaged within the virial radii, are shown in
Figure~\ref{fig:betax180}.

The prediction from the isotropic NFW model with concentrations given
by Equation~\ref{eq:xm180} is shown by the solid line.  It can be seen
that the theoretical curve mimics the data in that it predicts that
low-mass clusters should have lower values of $\beta_{X,180}$ than
high-mass ones.  However the mean measured values of $\beta_{X,180}$
lie about 0.1 below the predictions.  Once again, this could be
reconciled by using higher values of $x_{180}$ or an anisotropic dark
matter velocity ellipsoid.

Although virial temperatures are notoriously difficult to measure,
there is empirical evidence that the observed, X-ray temperatures of
the intracluster medium are greater than the virial temperatures
(e.g.~Edge \& Stewart 1991a,b; Bahcall \& Lubin 1994).  The latter
give a value of $\beta_{X,180}=0.94$ for the highest-mass clusters,
which is not too dissimilar from the values that we find in
Figure~\ref{fig:betax180}.  However, the results of the next section
suggest values that are lower than this.

\subsection{X-ray temperature-mass relations within a fixed overdensity}
\label{sec:txmvir}

\begin{figure}
  \begin{center} \psfig{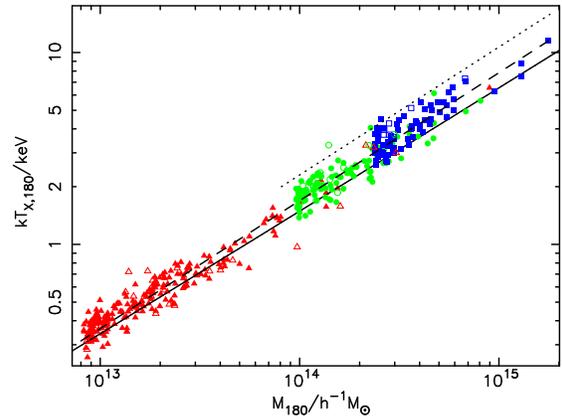}
    \caption{The X-ray temperature versus mass within the virial
    radius.  The dashed line shows the best-fit power law; the dotted
    line shows the best observational results from Horner, Mushotzky
    \& Scharf (1999) and the solid line shows the prediction from the
    isotropic NFW model.}  \label{fig:txm180} \end{center}
\end{figure}

The relationship between X-ray temperature and mass within the virial radius
is shown in Figure~\ref{fig:txm180}.  The dashed line shows the
best-fit relation from Evrard, Metzler \& Navarro (1996, hereafter EMN),
\begin{equation}
k\,T_{X,180} \approx 7.85\,
              \left(m_{180}\over10^{15}h^{-1}\Msun\right)^{0.67\pm0.02}\keV.
\label{eq:txm180}
\end{equation}
They extracted 58 clusters in
the temperature range 1--10\,keV, from three different sets of
cosmological simulations with a variety of cosmological parameters.
We have used the information given in their paper to interpolate their
results to an overdensity of 180---our results are in good agreement.

The temperature normalisation is slightly higher than for the
dynamical temperature-mass relation and corresponds to
$x_\delta\tilde{T}_\delta/\tilde{m}_\delta=0.49$.  The prediction from
Equation~\ref{eq:nfwtmvir} using values of $x_{180}$ taken from
Equation~\ref{eq:xm180} is shown by the solid line on the Figure.  Once
again it gives temperatures that are slightly too low.

The most extensive observational investigation of the $T_X$-$M_{\rm
180}$ relation is by Horner, Mushotzky \& Scharf (1999) who used many
different ways to determine the mass within the virial radius.  Their
preferred measure, based on X-ray emissivity and temperature profiles,
gives $T_X\propto M_{\rm 180}^{2/3}$, while those obtained from the
isothermal $\beta$-model and X-ray surface brightness deprojections
are flatter, $T_X \propto M_{\rm 180}^{0.5-0.56}$.  They suggest that
the slopes of these latter two measures are too low because they
assume a dark matter density profile of ${\rho_{\rm dark}\propto
r^{-2}}$ whilst observations and simulations suggest that it is
steeper, ${\rho_{\rm dark}\propto r^{-2.4}}$.

The preferred relation from Horner, Mushotzky \& Scharf (1999) is
shown as a dotted line in Figure~\ref{fig:txm180}.  From this it is
clear that either the measured cluster masses are too low, or more
likely dissipationless simulations predict X-ray temperatures that
are smaller than the observed values.  If we combine the results from
Figures~\ref{fig:betax180} and \ref{fig:txm180} then this argues that
observed high-mass clusters have values of $\beta_{X,180}\approx2/3$.

This high value of $\beta_{X,180}$ might be thought to provide evidence
for heating of the intracluster medium, but in fact radiative cooling
can have the same effect! Pearce \etal~(2000) have shown that the
removal of low-entropy gas by cooling in the core of the cluster can
raise the X-ray temperature of the cluster by 20 to 40 percent.  This
cooling is relatively more important in low-mass clusters and so would
have the effect of flattening the $T_X$-$M$ relation slightly.  

Mohr, Mathiesen \& Evrard (1999) find in a sample of 45 clusters that
the ICM is more extended than the dark-matter and that
$T_X\approxpropto M_{\rm gas}^{1/2}$.  In our simulations, we also
find that the gas is more extended than the dark matter, due to its
higher specific energy.  However, the baryon fraction within the
virial radius is approximately 0.85, independent of mass, and so
obtain a temperature-gas mass relation that parallels the one for the
total mass, $T_X\approxpropto M_{\rm gas}^{0.67}$.  Physical processes
such as heating or radiative cooling are once again required (and act
in the correct sense) to reconcile the observations and simulations.

\begin{figure}
  \begin{center} \psfig{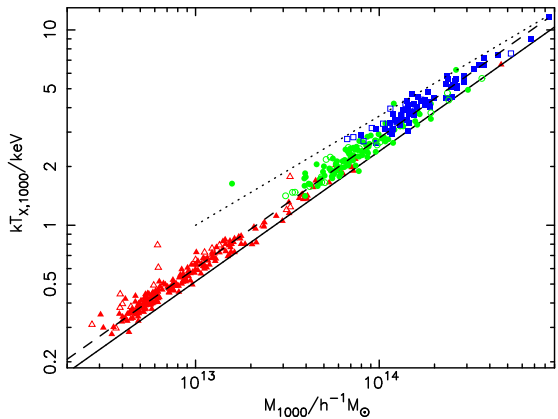}
    \caption{The X-ray temperature versus mass within a spherical
    shell enclosing an overdensity of 1000.  The dashed line shows the
    best-fit power law, the dotted line shows the observational
    results from Nevalainen, Markevitch \& Forman (2000), and the
    solid line shows the prediction from the isotropic NFW model.}
    \label{fig:txm1000} \end{center}
\end{figure}

In Figure~\ref{fig:txm1000}, we show a similar plot to
Figure~\ref{fig:txm180}, but for the X-ray temperature-mass relation
within an overdensity contour of 1000.
Once again, the isotropic NFW model slightly underpredicts the X-ray
temperatures (but agrees with the results of EMN).  The best-fit power
law is consistent with the self-similar prediction
\begin{equation}
k\,T_{X,1000} \approx 13.0\,
              \left(m_{1000}\over10^{15}h^{-1}\Msun\right)^{0.67}\keV.
\label{eq:txm1000}
\end{equation}

The dotted line shows the observational results from Nevalainen,
Markevitch \& Forman (2000).  The observed temperatures are again
higher than the predictions from this non-radiative model but in a way
that is now mass-dependent---this is consistent with our expectation
that the effects of cooling and/or heating would be greater in
lower-mass clusters.

\subsection{The X-ray temperature-virial radius relation}
\label{sec:txrvir}

From an observational point of view, it is much easier to measure the
virial radius rather than the virial mass, and so it is surprising
that more authors do not report the former.  There is no extra
information here, however, as the two are simply related and so it is
possible to rewrite Equations~\ref{eq:txm180}~and \ref{eq:txm1000} in
terms of virial radius and X-ray temperature:
\begin{equation}
r_{180} \approx
1.9\,\left(k\,T_{X,180}\over10\,\mbox{keV}\right)^{1\over2}\,h^{-1}\mbox{Mpc},
\label{eq:rtx180}
\end{equation}
\begin{equation}
r_{1000} \approx
0.85\,\left(k\,T_{X,1000}\over10\,\mbox{keV}\right)^{1\over2}
\,h^{-1}\mbox{Mpc}.
\label{eq:rtx1000}
\end{equation}
These follow with the self-similar form in that they have slopes of
${1\over2}$.  $r_{180}$ agrees with the results of EMN; $r_{1000}$ has a
slightly lower normalisation.

We note that the only observational measurement of $r_{1000}$ by Vikhlinin,
Forman \& Jones (1999) is at first sight in gross disagreement with our
prediction.  They have 
\begin{equation}
r_{1000} \approx
0.415\,\left(k\,T_{X,1000}\over10\,\mbox{keV}\right)^{0.57}
\,h^{-1}\mbox{Mpc}
\label{eq:rtxvfj99}
\end{equation}
which gives a value at 5\,keV of 1.04\,$h^{-1}$Mpc---far in excess of
the simulation results.  The reason for this discrepancy is that
Vikhlinin~\etal\ have defined overdensity with respect to a mean
baryon abundance of $\Omega_bh^2=0.010$ which is much lower than
either recent determination from primordial nucleosynthesis
($\Omega_bh^2=0.019$, Tytler \etal~2000) or from X-ray measurements of the
baryon fraction ($\Omega_bh^2=0.0594h^{1/2}$, Ettori \& Fabian 1999).
We can turn the argument around and ask what baryon abundance would
make our results compatible with Vikhlinin~\etal.  The radii they
quote enclose a baryon overdensity of about 230 in our models.  Hence
we require
\begin{equation}
\Omega_bh^2\approx0.010\,{1000\over230}\approx0.043,
\label{eq:omegab}
\end{equation}
This agrees with the value from Etorri \& Fabian for $h=0.54$.

\subsection{Temperature-mass relations within the Abell radius}
\label{sec:txmabell}

\begin{figure}
  \begin{center} \psfig{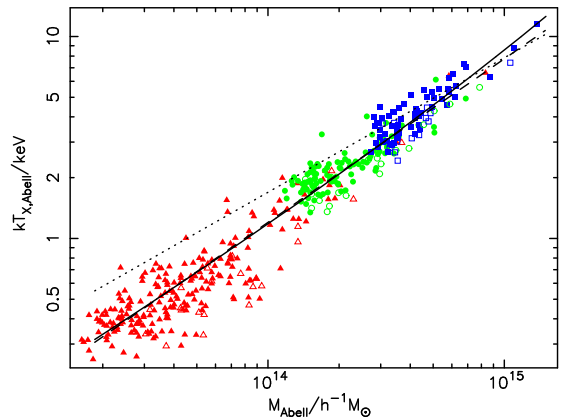}
    \caption{X-ray temperature versus mass, for properties averaged
    within the Abell radius, for all the clusters.  The dashed line shows
    the best-fit power law.  The dotted line is the relation from
    Equation~\ref{eq:txm180} for properties averaged within the virial
    radius.  The solid line shows the corrected relation as described
    in the text.}  \label{fig:txmabell} \end{center}
\end{figure}

If we measure cluster properties within a fixed radius, rather than a
fixed overdensity, then the deviation from the self-similar scaling
relations can be quite large.  This is shown in
Figure~\ref{fig:txmabell} where we plot, for within the Abell radius,
emission-weighted temperature versus mass.  The best-fit power law,
shown as the dashed line, has a slope of 0.81.

The reason for the steeper slope is that, for low-mass clusters, the
Abell radius is greater than the virial radius and so we are averaging
properties over a larger volume than before. This has the effect of
lowering the X-ray temperature slightly (because the X-ray temperature
is heavily weighted by emission from the centre of the cluster this
effect is small), but greatly increasing the mass.  For high-mass clusters,
however, the virial radii are similar to the Abell radii and so there
is no change.

Because the X-ray temperatures are almost unaltered, the
mass-temperature relation can be estimated by simply multiplying the
right-hand-side of Equation~\ref{eq:txm180} by the ratio of the mass
within an Abell radius to that within the virial radius, $M_{\rm
Abell}/M_{180}$.  We have done this, using Equation~\ref{eq:rtx180}
to estimate the virial radii and assuming that the mass profile
follows the NFW model.   The result, shown as the solid line in
Figure~\ref{fig:txmabell}, closely follows the data points.

\begin{figure}
  \begin{center} \psfig{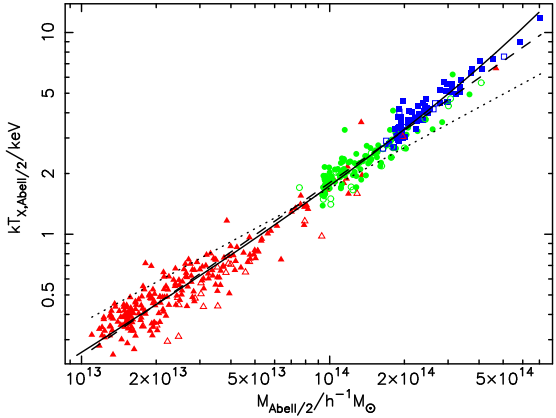}
    \caption{X-ray temperature versus mass, for properties averaged
    within half the Abell radius, for all the clusters.  The dashed line shows
    the best-fit power law.  The dotted line is the relation from
    Equation~\ref{eq:txm180} for properties averaged within the virial
    radius.  The solid line shows the corrected relation as described
    in the text.}  \label{fig:txmabhalf} \end{center}
\end{figure}

Figure~\ref{fig:txmabhalf}  is the same as for
Figure~\ref{fig:txmabell}, but for properties averaged within half an
Abell radius.  Once again the best-fit power law has a slope, 0.86,
that is steeper than the self-similar relation, and the corrected NFW
model provides an excellent fit to the data.

%% file: discussion.tex
\section{Discussion}
\label{sec:discuss}

\subsection{Comparison with previous simulations}
\label{sec:compsim}

Previous authors have also looked at scaling relations in simulations
of clusters without radiative cooling.  All agree that the gas is more
extended than the dark matter, but disagree on the details of the
$T_X$-$M$ relation.  The studies split into two types: those that
simulate a small number of clusters in detail, and those that, like
us, look at a large number of clusters at moderate resolution.

Navarro, Frenk \& White (1995) simulate only 6 clusters, extracted
from a low-resolution simulation, but choose these to span a factor of
thirty in mass and adjust the particle mass so that each cluster
contains several thousand particles within the virial radius.  They
find a $T_X$-$M$ relation in agreement with the virial relation and
values of $\beta$ of about unity (where their $\beta$, as ours,
includes a contribution from unthermalised motions in the gas) with no
dependence upon mass.  

EMN have taken the Navarro, Frenk \& White results and compared them
with two other sets of simulations of isolated clusters from Metzler
(1995) and Mohr \etal~(1995).  These span a range of cosmologies and
use two distinct codes, yet yield very similar results.  It should be
noted that EMN do not quote results for an overdensity of 180, but
they give sufficient information in their paper to allow one to
interpolate from their preferred overdensity of 500.  Once we do so,
we find a $T_X$-$M$ relation essentially identical to ours.

Eke, Navarro \& Frenk (1998) perform a similar study for 10 clusters
in a $\Lambda$-CDM cosmology, ranging over a factor of only three in
mass.  Their results are consistent with the scaling relations but
cover too small a dynamic range to provide a strong constraint.

By contrast, Bryan \& Norman (1998) use a smaller box and simulate a
large number of clusters at relatively poor resolution.  They find
that $\beta$ is an increasing function of cluster mass which suggests
that numerical heating is having an effect.  They also find a
$T_X$-$M$ relation that is flatter than the self-similar one, but only
for an open cosmology.

Yoshikawa, Jing \& Suto (2000) have undertaken similar simulations to
ourselves, but with a poorer spatial resolution and covering a smaller
mass-range.  They too find the expected slope of ${2\over3}$ for the
$T_X$-$M$ relation.

\subsection{Beyond the isotropic NFW model}
\label{sec:nfwani}

It can be seen from Figures~\ref{fig:tm180} and \ref{fig:txm180} that
the isotropic NFW model underpredicts the virial temperature slightly
and the X-ray temperature quite a lot.

One possible explanation for the low virial temperatures is that the
dark matter velocity ellipsoid is not isotropic, as radial orbits are
less efficient at supporting the particles within a given potential
than transverse ones.


We have checked that a ratio of the transverse to radial velocity
dispersion of $\sigma_t=0.9\,\sigma_r$ is sufficient to bring the
dashed and solid lines in Figures~\ref{fig:tm180} and
\ref{fig:txm180} into close agreement.  
We choose not to report on this in detail, however, because we believe
that it overly-complicates the simple model that we are trying to
test.  It should be remembered that few of our clusters are free from
some form of substructure and this too would invalidate the simple NFW
model.

We prefer to conclude that the isotropic NFW model does pretty well, whilst
bearing in mind that it underestimates the temperatures slightly.

%% file: conclusions.tex
\section{Conclusions}
\label{sec:conclude}

We extract clusters, covering more than 2 decades in mass, from
three simulations of the $\tau$CDM cosmology.  This represents the
largest, uniform catalogue of simulated clusters ever produced.

We fit the circular-speed profiles of the clusters with models
corresponding to spherically-symmetric density profiles of the form
\begin{equation}
\rho={\rho_0\over x\,(1+x)^s},
\end{equation}
where $s=2$ corresponds to the renowned NFW profile.

We find that the scatter in the best-fit values of $s$ is large, with
under a quarter of the clusters being accurately fit by the NFW
model.  The others either have steeper outer slopes in their density
profiles, show a sharp break in their density profiles that cannot be
fit by the above form, or have significant substructure.  However, the
mean, smooth, low-mass cluster does have $s\approx2$, whereas the
equivalent high-mass cluster has a steeper profile with $s\approx2.8$.

When we force $s=2$, then the cluster concentrations show a large
scatter, but the median concentration declines slightly with mass.
This is driven primarily by an anti-correlation between concentration
and substructure, with substructure being more prevalent in high-mass
clusters.

We investigate how well this median NFW model predicts the cluster
temperature-mass scaling relations, by deriving theoretical
relationships between both dynamical and emission-weighted, X-ray
temperature and mass in the isotropic NFW model, as a function of
cluster concentration.
The virial temperature-mass relation, averaged within a spherical
region enclosing an overdensity of 180, closely mimics
the self-similar form and lies a few per cent above the NFW prediction.

The gas in our clusters is hotter than the dark matter but we do not
attach much significance to this: there will be a small amount of
true and numerical heating due to heating by clumps and individual
particles of dark matter.  More importantly, we have chosen in this
paper to neglect the effects of radiative cooling and heating
associated with metal enrichment of the intracluster medium.

When measured within spheres enclosing a fixed overdensity, the X-ray
temperature versus virial mass relation has a slope of approximately
0.67, in agreement with the self-similar prediction and with previous
work (but extending over a greater mass-range).  The normalisation is
lower than that of the observations.  This is almost certainly because
we have chosen to neglect cooling and/or heating.

The radius-temperature relation is in agreement with the measured
value of Vikhlinin, Forman \& Jones (1999) provided that the baryon
fraction is large, as indicated by Ettori \& Fabian (1999).

When averaged within an Abell radius (or half an Abell radius), the
X-ray temperature versus mass relation is steeper, with a slope of
0.81 (or 0.86).  This is because the Abell radius is a greater
multiple of the virial radius in low-mass clusters as compared to
high-mass ones and so increases their measured mass.

We have not commented on the X-ray luminosity in this paper, because
our simulations are not able to fully resolve the X-ray emission in
the cluster cores.  This core gas has a cooling time that is anyway
less than the age of the Universe and so cannot be correctly modelled
using a non-radiative simulation.  We note that the results of
Pearce~\etal (2000) suggest that radiative cooling will act so as to
\emph{raise} the temperature of the intracluster medium and so tend to
bring our simulated clusters into agreement with the observations.

In future papers we will contrast the dynamics of the gas and dark
matter in clusters, consider the effect of heating and cooling
processes, and compare the results from different cosmological models.

%% file: references.tex
\bsp